\documentclass[twoside,12pt]{article}
\usepackage{graphicx}
\begin{document}
\centerline{\bf Simulation of never changed opinions in Sznajd}
\centerline{\bf consensus model using multi-spin coding}

\bigskip
Dietrich Stauffer$^1$ and Paulo Murilo C. de Oliveira$^2$

\bigskip
Laboratoire PMMH, Ecole Sup\'erieure de Physique et Chimie Industrielle, 10
rue Vauquelin, F-75231 Paris, Euroland

\medskip
\noindent
$^1$ Visiting from Institute for Theoretical Physics, Cologne University, 
D-50923 K\"oln, Euroland; stauffer@thp.uni-koeln.de

\noindent
$^2$ Visiting from  
Instituto de F\'{\i}sica, Universidade Federal Fluminense; Av. Litor\^{a}nea 
s/n, Boa Viagem, Niter\'{o}i 24210-340, RJ, Brazil; pmco@if.uff.br
\bigskip

Abstract: 
The density of never changed opinions during the Sznajd consensus-finding 
process decays with time $t$ as $1/t^\theta$. We find
$\theta \simeq 3/8$ for a chain, compatible with the exact Ising result of 
Derrida et al. In higher dimensions, however, the exponent differs from the
Ising $\theta$. With simultaneous updating of sublattices instead of the
usual random sequential updating, the number of persistent opinions decays
roughly exponentially. Some of the simulations used multi-spin coding.
\medskip


\bigskip

In the Ising and Potts models, the persistent spins [1] are those which from
the beginning of a (zero-temperature) Monte Carlo simulation have never been 
flipped. In the thermodynamic limit, their number
decreases with time $t$ asymptotically as $1/t^\theta$ with $\theta = 3/8$
exactly on the Ising chain [2], while higher dimensions were investigated 
numerically [3] giving $\theta \simeq 0.2$ on the square lattice. Also
the more general Potts model was investigated [1,2,3]. Now we
simulate the analogous number in a $d$-dimensional Sznajd model of 
consensus-finding [4] with up to 49 million sites and $1 \le d \le 4$, for the 
case of just two possible opinions, the equivalent of spin 1/2 Ising sites.

\begin{figure}[hbt]
\begin{center}
\includegraphics[angle=-90,scale=0.50]{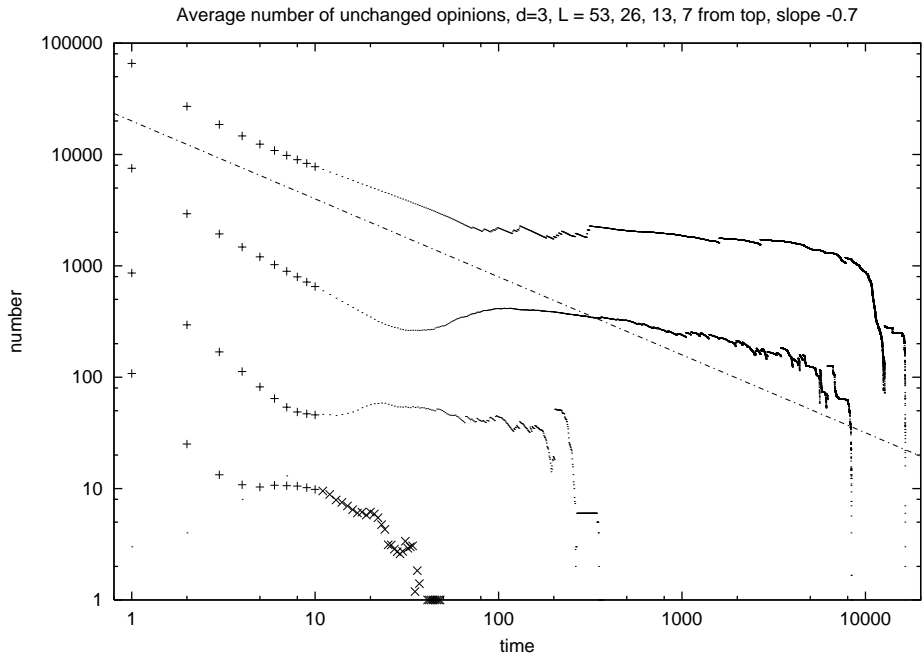}
\end{center}
\caption{
Log-log plot of the number of persistent people, $P(t)-P(\infty)$, 
versus time, for one, $10^3, \, 10^3, \, 10^4$ simple cubic lattices of size 
$L \times L \times L$ with $L = 53$, 26, 13, 7. Only the intermediate  times
before the plateau and final decay to zero are used to estimate the
exponent $\theta \simeq 0.7$, indicated here by the straight line.
}
\end{figure}

\begin{figure}[hbt]
\begin{center}
\includegraphics[angle=-90,scale=0.50]{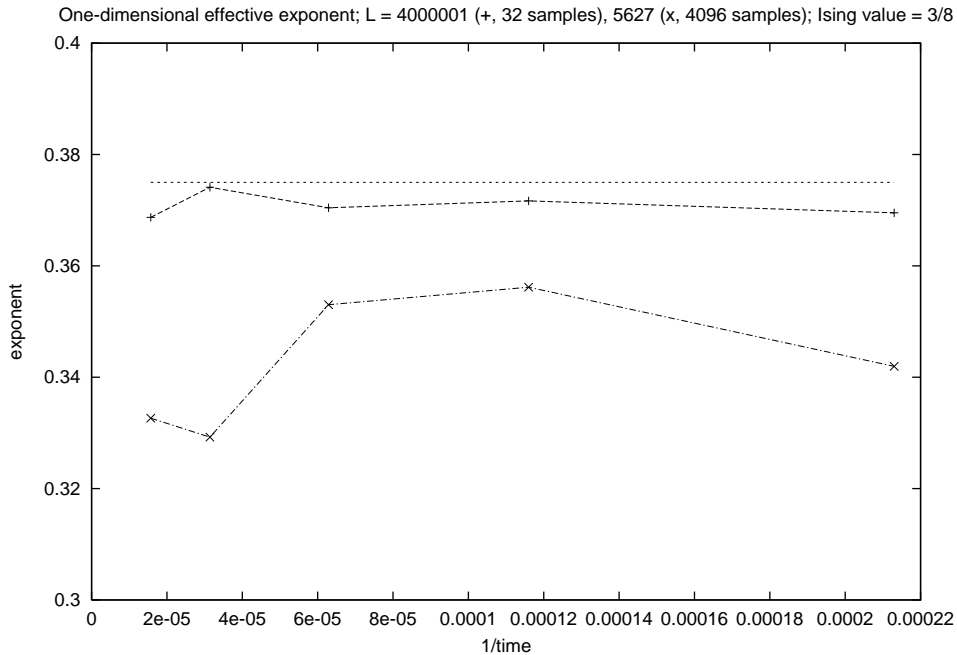}
\end{center}
\caption{
Effective exponents in one dimension, approaching perhaps the Ising
value 3/8 (horizontal line) for long times and large lattices.
}
\end{figure}

\begin{figure}[hbt]
\begin{center}
\includegraphics[angle=-90,scale=0.50]{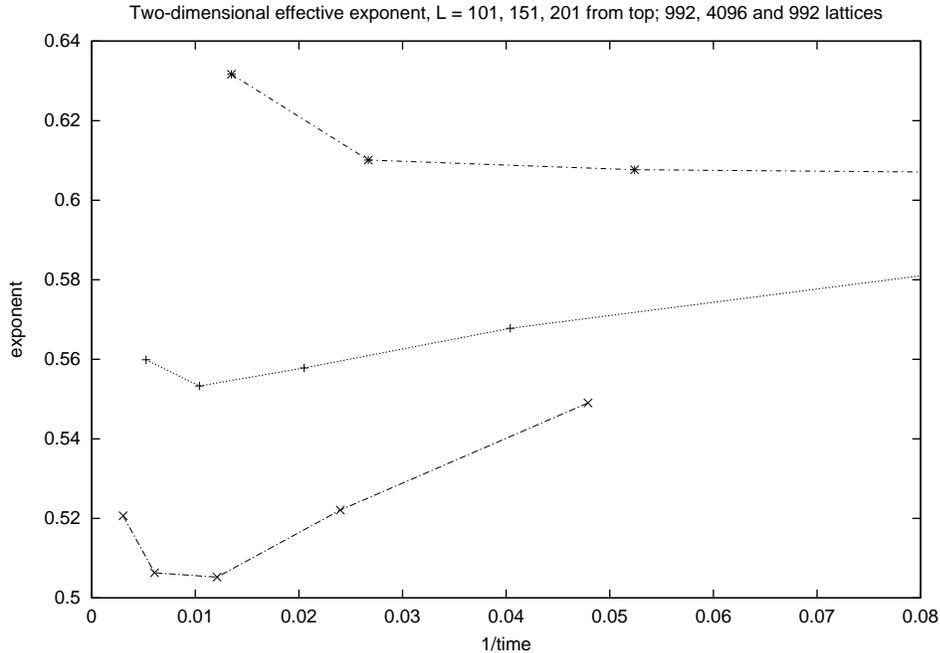}
\end{center}
\caption{
As Fig.2, but for square lattices, using intermediate times only, before any
final consensus was found.
}
\end{figure}

\begin{figure}[hbt]
\begin{center}
\includegraphics[angle=-90,scale=0.50]{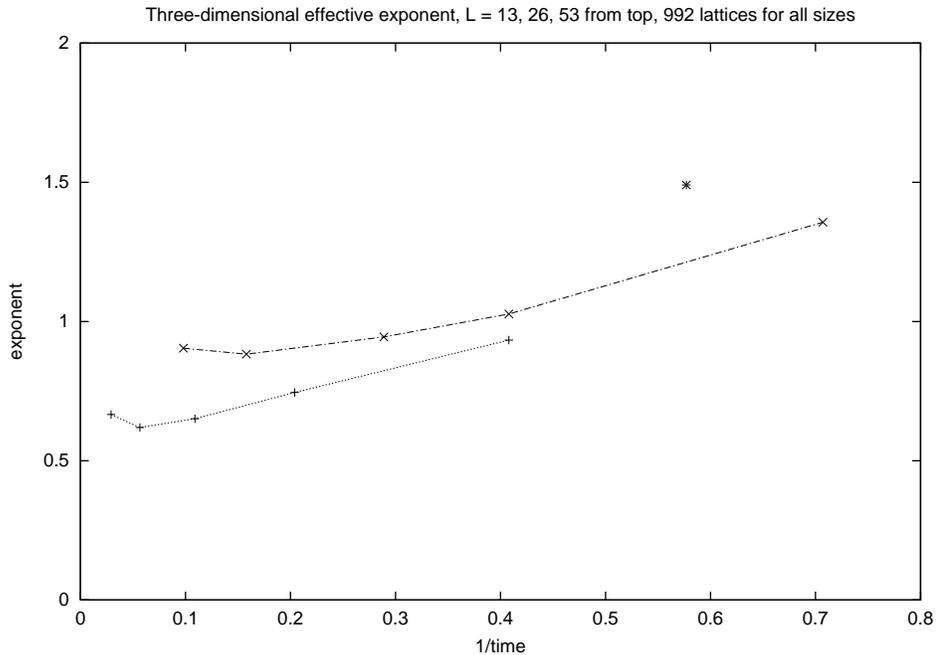}
\end{center}
\caption{
As Fig.3 but for simple cubic lattices. 
}
\end{figure}

\begin{figure}[hbt]
\begin{center}
\includegraphics[angle=-90,scale=0.50]{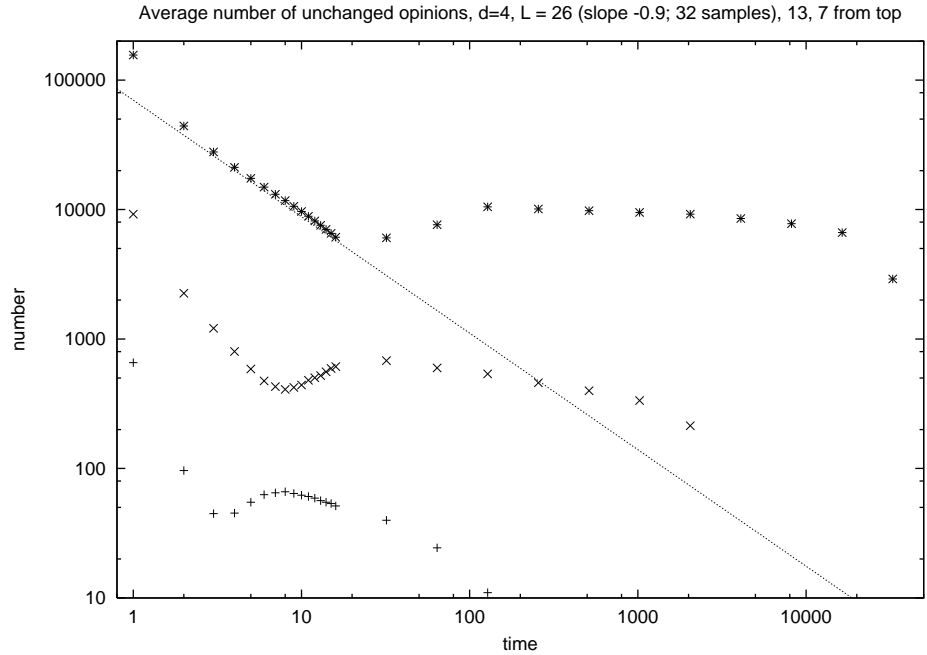}
\end{center}
\caption{
Log-log plot of $P(t)-P(\infty)$ for hypercubic lattices in four 
dimensions averaged over $10^3$ samples (32 samples only for $L = 26$).
}
\end{figure}

\begin{figure}[hbt]
\begin{center}
\includegraphics[angle=-90,scale=0.50]{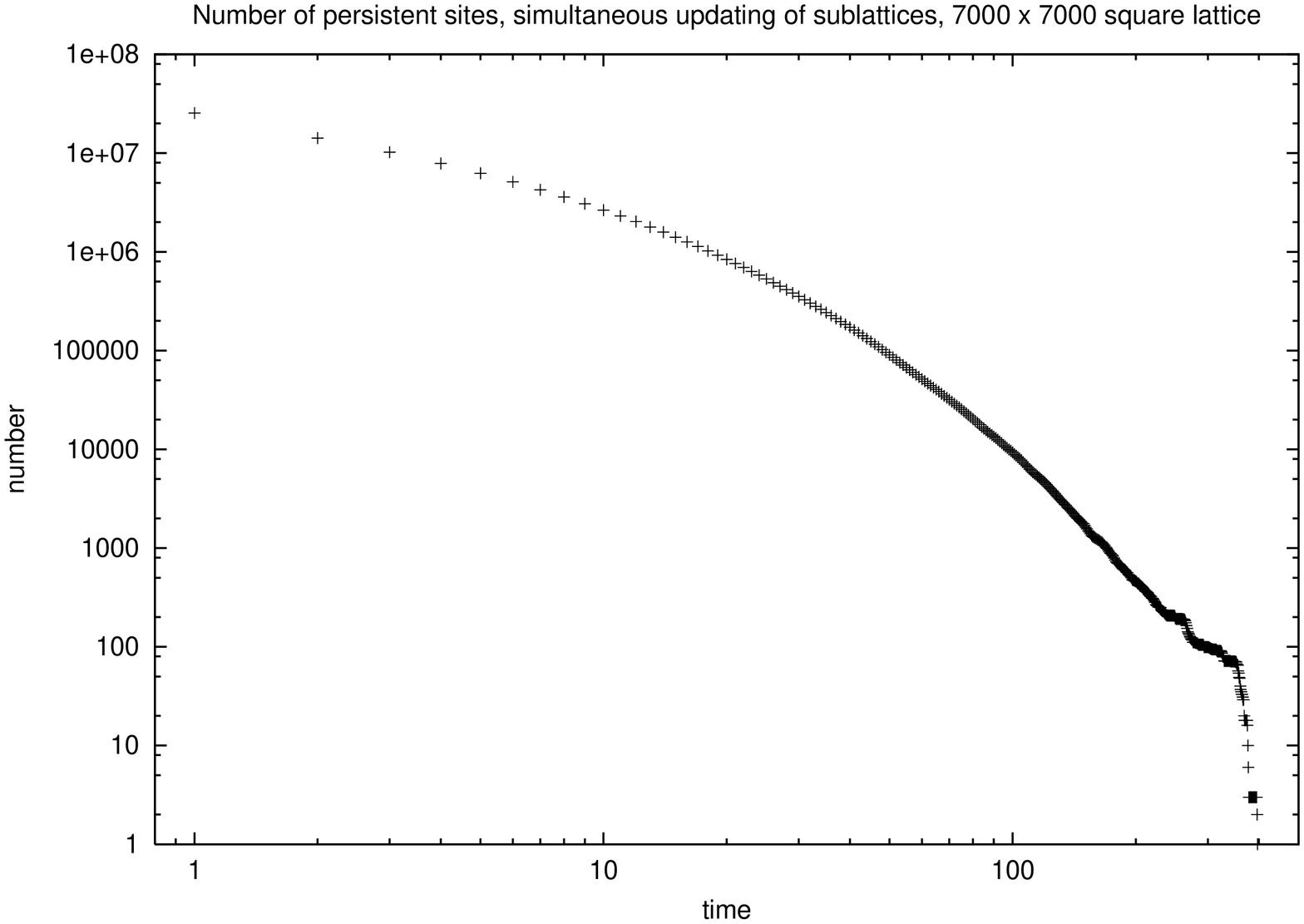}
\end{center}
\caption{
Log-log plot of $P(t)$ for $7000 \times 7000$ square lattice
with unfrustrated simultaneous updating of sublattices. 
}
\end{figure}
In this Sznajd model (see [5] for a review) two opposing opinions are 
initially distributed randomly with equal probability over the $L^d$ 
``people'' of a hypercubic lattice. Then, each randomly selected pair of 
nearest neighbours convinces its $4d-2$ nearest neighbours of the pair
opinion if the pair shares the same opinion; otherwise, the neighbour opinions
are not affected. One time step means that on average every lattice site is
selected once as the first member of the pair. (We will mention below the
different results if this random sequential updating is replaced by 
simultaneous updating.) The Sznajd model is one of several recent 
consensus-finding models [6] and follows a long tradition of social 
studies using computer simulation and/or statistical physics [7]. If we
wait sufficiently long for large systems, always a consensus is found:
Everybody has the same opinion and the whole system has reached a fixed
point. 

Alternatively, independently of social interpretations, this model can be 
understood as a variant of the traditional kinetic Ising model: instead of a 
central site being influenced by its neighbourhood, the neighbourhood itself
is updated according to the states of the central spins.

We check for the number $P(t)$ of ``persistent'' sites who have not yet
changed their spins in this Sznajd consensus process. (All our sites are
equivalent, in contrast to Schneider's modification [8] where some sites are
initially selected as permanent opponents.) We find that usually a consensus is
found before everybody had changed opinion; i.e. $P(\infty) > 0$. Thus the
exponent $\theta$ has to be determined from intermediate times where 
$P(0) \gg P(t) \gg P(\infty),$ or from $P(t) - P(\infty)$. Fig.1 shows that
this latter quantity has a complicated behaviour, and again only intermediate
times are used to find $\theta$. In one dimension, $P(\infty)$ is relatively
small and the resulting systematic deviations disturb less.

Figs.2 to 4 show for $d =$ 1, 2 and 3 the effective exponents $\theta$ analyzed
by least square fits over five suitable time intervals $t_n < t < t_{n+1}$ with
$t_{n+1} = 2 t_n$. For $d>1$, only the times until the first of the (typically 
1000) samples reached a consensus were used and averaged over. We conclude that
$\theta \simeq $ 3/8 in one dimension,  0.5 in two, and the same or somewhat 
higher in three dimensions. Fig.5 
shows that four dimensions is difficult to analyze though maybe $\theta \simeq
0.9$. Our one-dimensional estimate is compatible with the Ising value
3/8, but for higher dimensions our $\theta(d)$ goes up while the Ising 
$\theta(d)$ went down for increasing $d$.

We also speeded up the simulations by storing 32 or 64
sites (belonging to 32 or 64 different samples) in each computer word,
using single-bit handling [9] known for Ising models as multi-spin coding.
The random selection of neighbour pairs was the same for all 32 or 64
samples. The C program is available from PMCO, the Fortran program from DS.

If during one time step all sites are updated simultaneously, with frustrated
sites not changing their opinion, then no consensus is found [10]. (Frustrated
are those sites which simultaneously are convinced to different opinions
by different neighbour pairs.) This frustration can be avoided by dividing
the lattice into sublattices, such that  no sites within one sublattice
can influence each other directly; we divided our lattices such that the 
distances between sites belonging to the same sublattice are at least five
lattice constants. (For nearest-neighbour Ising models, the two sublattices
of a chessboard suffice on the square lattice, while we used 25 
interpenetrating sublattices for the square Sznajd model.) With this 
simultaneous updating of sublattices, frustration is avoided, a consensus is
always found, but $P(t)$ no longer decays as a power law, Fig.6: criticality 
seems lost. Also,
this version no longer shows the phase transition of the usual square Sznajd
model, when the ratio of the fractions of the two initial opinions is varied 
away from unity. Thus, the simultaneous updating is not merely a possible
acceleration of the dynamic process. Also the correlations between spins 
behave differently, being affected by the simultaneous updating of spins 
far apart from each other.

In summary, the Sznajd model is Ising-like in one dimension but not in higher
dimensions for the persistence exponent $\theta$.

We thank B. Derrida for a discussion, and the J\"ulich supercomputer center 
for time on their Cray-T3E.

\bigskip
\parindent 0pt

[1] B. Derrida, A.J. Bray, C. Godreche, J. Phys. A 27, 357 (1994).

[2] B.Derrida, V. Hakim, and V. Pasquier, Phys. Rev. Lett. 75, 751 (1995).

[3] D. Stauffer, J. Phys. A 27, 5029 (1994); B. Derrida, P.M.C. de Oliveira
and D. Stauffer,  Physica A 224, 604 (1996).

[4] K. Sznajd-Weron and J. Sznajd, Int. J. Mod. Phys. C 11, 1157 (2000).

[5] D. Stauffer, Journal of Artificial Societies and Social Simulation,
vol. 5, issue 1, paper 4 (Feb. 2002), available at jasss.soc.surrey.ac.uk

[6] G. Deffuant, D. Neau, F. Amblard and G. Weisbuch,
Adv. Complex Syst. 3, 87 (2000);  G. Weisbuch, G. Deffuant, F. Amblard, and
J.-P. Nadal, Complexity 7, 55 (2002) (cond-mat/0111494);
R. Hegselmann and M. Krause, Journal of Artificial Societies
and Social Simulation 5, issue 3, paper 2 (2002) (jasss.soc.surrey.ac.uk);
see also J.C. Dittner, Nonlinear Analysis 47, 4615 (2001).

[7] T.C. Schelling, J. Mathematical Sociology 1, 143 6(1971);
 E. Callen and D. Shapero, Phys. Today July 1974, p. 23;
W. Weidlich, {\it Sociodynamics; A Systematic Approach
to Mathematical Modelling in the Social Sciences}. Harwood Academic
Publishers, 2000; S. Galam, Y. Gefen and Y. Shapir, J. Mathematical
Sociology 9, 13 (1982);
F. Schweitzer, (ed.) {\it  Self-Organization of Complex Structures:
From Individual to Collective Dynamics}, Gordon and Breach, Amsterdam 1997;
J.A. Ho{\l}yst, K. Kacperski and F. Schweitzer, in {\it
Annual Reviews of Computational Physics} IX, p.275, World Scientific,
Singapore 2001;
F. Schweitzer and K.G. Treutzsch (eds.), SocioPhysics, conference abstracts,
Bielefeld (Germany), June 2002, ais.gmd.de/~frank/sociophysics; S.Moss de
Oliveira, P.M.C. de Oliveira and D. Stauffer, {\it Evolution, Money, War and
Computers}, chapter 5, Teubner, Leipzig Stuttgart ISBN 3-0519-00279-5 (1999).

[8] J. Schneider, poster at Bielefeld conference [7].

[9] G. Bhanot et al., J. Stat. Phys. 44, 985 (1986); 
P.M.C. de Oliveira, {\it Computing Boolean Statistical Models}, World 
Scientific, Singapore London New York, ISBN 981-02-238-5 (1991); 

[10] D. Stauffer, submitted to J. Mathematical Sociology.
\end{document}